\documentstyle[aps,amssymb,epsfig,multicol]{revtex}
\def\be{\begin{equation}}
\def\ee{\end{equation}}
\def\bq{\begin{eqnarray}}
\def\eq{\end{eqnarray}}
\def\bm{\begin{multicols}{2}}
\def\em{\end{multicols}}

 1

\begin{document}

\draft
\title{Exact thermodynamics of an Extended Hubbard Model
of single and paired carriers in competition}

\author{Fabrizio Dolcini and Arianna Montorsi
\footnote{e-mail: fdolcini@athena.polito.it \, ;
\, montorsi@athena.polito.it}}
\address{Dipartimento di Fisica and Unit\`a INFM, Politecnico
di Torino, I-10129 Torino, Italy}
\date{\today}
\maketitle
\begin{abstract}
By exploiting the technique of Sutherland's species, introduced in
\cite{DOMO-RC}, we derive the exact spectrum and partition
function of a 1D extended Hubbard model. The model describes a
competition between dynamics of single carriers and short-radius
pairs, as a function of on-site Coulomb repulsion ($U$) and
filling ($\rho$). We provide the temperature dependence of
chemical potential, compressibility, local magnetic moment, and
specific heat. In particular the latter turns out to exhibit two
peaks, both related to `charge' degrees of freedom. Their origin
and behavior are analyzed in terms of kinetic and potential
energy, both across the metal-insulator transition point and in
the strong coupling regime. 
\end{abstract}

\begin{multicols}{2}
\section{Introduction}
In condensed matter, electron systems in regimes of high
correlation are known to be suitably modeled by the Hubbard
Hamiltonian\cite{HUB} and its
generalizations\cite{GEB-bk,highTc,EHM-ex,AAS,POL,PKH,EHM-vr}. For
such models, the finite temperature properties are the ultimate
results which theoretical investigations (numerical or
analytical) aim to reach, in view of comparisons to experimental
data. Indeed some observables exhibit intriguing features as a
function of the temperature, which deserve an accurate
interpretation.
\\In particular, the thermodynamics of the standard Hubbard model has
been widely investigated. In D=1 this was done by different exact
approaches: in \cite{TAK,KUO} and in \cite{JKS} for the usual
case of nearest-neighbors hopping, while in \cite{GEB} for the
case of long-range hopping. In dimension greater than one recent
results were obtained by exact diagonalization on small
clusters\cite{PINCUS,SCHU} and numerical
investigations\cite{PASC,DUFFY}, whereas the case D=$\infty$ has
been examined in \cite{GEO} by iterated perturbation theory.
\\All the results show interesting behaviors as a function of
temperature, with varying the filling and the Coulomb repulsion.
This is the case for instance for the specific heat, where a
double peak structure as well as the appearance of quasi
universal crossing points were found, which features were already
noticed in some experimental data\cite{HEAVY1,HEAVY2}. In the
strong coupling regime the presence of a two-peak structure is
usually related to the so called `spin' and `charge' degrees of
freedom. Numerical results in one \cite{KUO} and two
dimensions\cite{PASC,DUFFY} show that, at least at half-filling,
such structure survives also
at moderate couplings.\\

Contrary to the ordinary Hubbard model, which has been approached
through several techniques, for the extended Hubbard models
most of finite-temperature results have been carried out by means of
mean-field theories\cite{POL}. In one dimension, however, it is
known that traditional approaches to many-body systems such as
mean-field or Fermi Liquid theories are either unreliable or
inapplicable. As a consequence, both numerical techniques (like
Density Matrix Renormalization Group\cite{DMRG}) and
non-conventional analytical approaches (like
bosonization\cite{BOS}) have to be supported by comparison with
exact solutions, whenever available; this is basically the reason
for the growing interest devoted to finite temperature {\it
exact} results. \\The main technique within exact approaches to
one dimensional systems is the Bethe Ansatz (BA), either in the
Coordinate\cite{CBA} or in the Algebraic\cite{ABA} formulation.
Such technique amounts to guessing for a given model eigenstates
of the form proposed by Bethe\cite{BETHE}, and in particular it
has been extensively applied to models of correlated electrons;
for instance, the BA equations for a wide class of integrable
extended Hubbard models \cite{DOMO-IJMB} have been recently
derived in \cite{AUS}. However, the actual solution of these
equations, {\it i.e.} the evaluation of the quantum numbers
characterizing the system (quasi momenta), is in general quite
difficult, and some hypothesis on their distribution (string
hypothesis\cite{string}) have typically to be conjectured. In
order to derive the complete solution and calculate thermodynamic
quantities, one is thus reduced to solve a system of infinitely
many coupled integral equations, which requires dramatic
numerical effort. More recently, a considerable progress has been
achieved through the alternative approach of Quantum Transfer
Matrix\cite{QTM}, which yields dealing with only a finite number
of coupled integral equations. This has been done for the
ordinary Hubbard model\cite{JKS}, for the $t-J$ model
\cite{JKS-tJ}, as well as for an extended Hubbard model with
bond-charge
interaction\cite{JKS-ch}.\\
Nevertheless, determining the actual properties of
a model at finite temperature for arbitrary parameters' values
remains in general a very hard task, even when the model is proved
to be integrable, and its ground state features are possibly
derived.

In the present paper we present the exact thermodynamics of a
one-dimensional Extended Hubbard Hamiltonian (described in sec.
II) whose exact analytical ground state properties were obtained
in \cite{DOMO-RC} by a technique different from BA. We called that
technique Sutherland Species (SS) technique, and here we show how
it can be exploited to derive explicitly the whole spectrum and
the partition function of the model (section III).  In Section IV
we calculate some thermodynamic quantities, namely the
chemical potential, the compressibility, the local magnetic
moment, and the specific heat. In particular in subsection D we
focus on the specific heat, which turns out to exhibit a two-peaks
structure. The origin of such structure and the differences with
respect to the standard Hubbard model are
discussed in~section~V.\\
\section{The model}
\begin{figure}
\epsfig{file=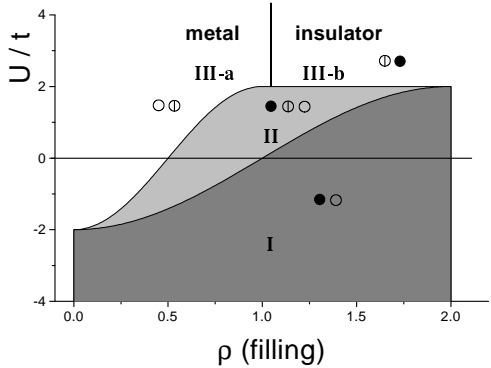,width=7cm,height=5.5cm,clip=}
\caption{Ground state phase diagram of the model (\ref{HAM}) for
$Y=t$, from [1]. Empty, bared, and full circles respectively
represent empty, singly occupied, and doubly occupied sites in the
ground state.} \label{ter_fig1}
\end{figure}

The Hamiltonian we are interested in reads:
\end{multicols}
\begin{eqnarray}
&\hat{\mathcal H}& = - t \sum_{\langle{i},{j}\rangle,\sigma }
(1-\hat{n}_{{i} \, -\sigma}) \,
 c_{{i} \, \sigma}^\dagger
c_{{j} \, \sigma}^{} (1-\hat{n}_{{j}\, -\sigma}) \, \,
+ \, \,  Y\sum_{\langle{i},{j}\rangle } c_{{i} \uparrow}^\dagger
c_{{i} \downarrow}^\dagger  c_{{j} \downarrow}^{} c_{{j}
\uparrow}^{} \, \, + \, \, U \sum_{i} \hat{n}_{{i} \uparrow}
\hat{n}_{{i} \downarrow}
\label{HAM}
\end{eqnarray}
\begin{multicols}{2}
\noindent Here $c_{{i}\, \sigma}^\dagger , c_{{i} \,\sigma}^{}
\,$ are fermionic creation and annihilation operators on a
one-dimensional chain with $L$ sites, $\sigma \in \{\uparrow,
\downarrow \}$ is the spin label, $\, \hat{n}_{{j} \,\sigma} =
c_{{j} \, \sigma}^\dagger c_{{j}\,\sigma}^{}$, and $\langle {i} ,
\, {j} \rangle$ stands for neighboring sites. The Fock space
$\mathsf{F}$ of the system is the product of the $L$
four-dimensional vector spaces $V_j$ related to each site $j$;
each $V_j$ is spanned by the basis $|\uparrow\rangle_j
,|\downarrow \rangle_j ,|0\rangle_j ,|\downarrow \uparrow
\rangle_j$, which we shall also denote in the following as
$|e_\alpha\rangle_j$, $\alpha=1,\ldots 4$ respectively.  We shall
adopt for the 1D lattice {\it open} boundary conditions; as usual,
these are not expected to affect the results in the
thermodynamic limit.
\\In the Hamiltonian (\ref{HAM}) the three terms (which will also
be denoted as ${\mathcal{H}}_t$, ${\mathcal{H}}_Y$ and
${\mathcal{H}}_U$) represent respectively the kinetics of single
carriers, the kinetics of paired carriers, and the on-site Coulomb
repulsion.
\\More explicitly, ${\mathcal{H}}_t$ describes the hopping of
single electrons towards empty sites. This term is thus
reminiscent of the so-called '$U=\infty$ Hubbard model'. An
important difference must be however highlighted: the latter
model reads ${\mathcal P} \sum_{\langle i, j \rangle \, , \sigma}
c_{{i} \sigma}^\dagger c_{{j}  \sigma}^{} \, {\mathcal P}$, where
${\mathcal P}=\prod_{i} (1-\hat{n}_{i \uparrow} \hat{n}_{i
\downarrow})$ projects the doubly occupied sites out of the
Hilbert space (which in that case is actually $3^L$-dimensional);
in contrast, the term ${\mathcal{H}}_t$ in (\ref{HAM}), although
not involving pairs, does not exclude their presence in the state
of the system\cite{NOTAUinf}.
\\The second term in (\ref{HAM}) is in contrast a kinetic term of
pairs only; it is worth stressing that the model deals with pairs
having a very short radius; in fact, while in models such as BCS
one has several pairs within a radius of the coherence length,
here the radius of a pair is thought of as small with respect to
the lattice constant, and is actually taken as zero. This kind of
term is also used in the so called Penson-Kolb-Hubbard model (see
\cite{PKH}), where one investigates the effects of the pair
dynamics without explicitly entering the microscopic mechanism
yielding their formation\cite{NOTE}. We also point out that the
first and the second term in (\ref{HAM}), though describing the
kinetic of different kind of carriers (single and pair
respectively), {\it do not} commute at all.
\\The third term is traditionally the most important term
for Hubbard-like models; indeed, according to Hubbard's picture,
it is the parameter that should drive the metal-insulator
transition in the $d$-transition metal compounds. Loosely
speaking, the ratio $U / t$ can be thought as proportional to the
inverse of the pressure applied on the sample: by increasing the
pressure one reduces the lattice spacing and thus makes the
hopping amplitude more relevant with respect to $U$.
\\The first two terms of the Hamiltonian are
in general competing: indeed ${\mathcal{H}}_t$ would favor
delocalized waves of single carriers, avoiding the formation of
pairs; ${\mathcal{H}}_Y$ lowers instead the energy when electrons
form tightly bound pairs moving along the chain. This competition
is in addition modulated by both the term in $U$ and the filling,
{\it i.e.} the density $\rho$ of electrons in the chain. This can
be seen by examining the case
\begin{equation}
Y=-t \quad . \label{Yeqt}
\end{equation}
Indeed for this value of the coupling constant the model has been
proved to be integrable\cite{DOMO-IJMB} and the exact ground
state phase diagram (reported in fig. \ref{ter_fig1}) has been
obtained in\cite{DOMO-RC}. Tuning $U$ and $\rho$ the model
exhibits interesting features; for instance, even when the value
of filling is $\rho < 1$ and at moderate ($U<2 t$) Coulomb
repulsion, it is energetically favorable for the system to form
pairs and let them move instead of having only singly occupied
sites.
\\In region I the ground state (GS) is made of only doubly occupied
and empty sites; in region II we have also singly occupied sites
(either $|\uparrow \rangle$ or $|\downarrow \rangle$). In region
III-a the GS is that of the $U=\infty$ Hubbard model and
is made of singly occupied sites (metal). In region III-b the GS
of the model reduces to that of the atomic limit of the Hubbard
model (insulator). At half-filling ($\rho=1$) a charge gap
$\Delta_c=U-2t$ opens for any $U \ge 2t$.
\\We wish to stress that, unlike many exactly solved electron
systems, the model (\ref{HAM}) is not particle-hole invariant:
indeed the first term breaks up the invariance; this leads to
the shape of the phase diagram shown in fig.\ref{ter_fig1}, which is
asymmetrical with respect to half-filling.
\section{Spectrum of the system}
In the following we shall assume $Y=-t$, since such relation
allows for the integrability, as observed above. In this case, the
Hamiltonian (\ref{HAM}) can be rewritten in the form \be
\hat{\mathcal H}= \sum_{\langle i,j \rangle} \hat{T}_{i,j} + U
\sum_i \hat{n}_{i \uparrow} \hat{n}_{i \downarrow} \label{HAM2}
\ee where $\hat{T}_{i,j}$ accounts for the first two interaction
terms in (\ref{HAM}). The term in $U$ is easily checked to commute
with $\sum_{\langle i,j \rangle} \hat{T}_{i,j}$. Due to the
condition (\ref{Yeqt}) $\hat{T}_{i,j}$, exhibits the structure of
a {\it generalized} off-diagonal permutator between Physical
Species (PS), which are the local vectors $|e_\alpha\rangle$'s. More
explicitly, while the ordinary off-diagonal permutator, when acting onto
$|e_\alpha \rangle_j \otimes |e_\beta \rangle_{j+1}$, returns
$|e_\beta\rangle_j \otimes |e_\alpha \rangle_{j+1}$ for {\it any}
$\alpha \neq \beta$ and zero for $\alpha=\beta$, a generalized
one makes the exchange or gives zero according to the specific
values of $\alpha$ and $\beta$. In our case, $\hat{T}_{i,j}$
permutes the PS of two neighboring sites only if one belongs to
the group $A$ and the other to the group $B$, where
\begin{equation}
A=|\uparrow\rangle,|\downarrow\rangle,|\downarrow \uparrow\rangle
\quad , \quad  B=|0\rangle \quad .
\end{equation}
In all the remaining cases $\hat{T}_{i,j}$ gives zero. The above
groups $A$ and $B$ of PS can be identified with the Sutherland Species
(SS) of the model (\ref{HAM})(see \cite{DOMO-RC}); the notion of SS is
strictly related to the structure of the Hamiltonian and not to that of
the underlying Hilbert space \cite{NOTAGP}.\\
In D=1 a generalized permutator between PS has the same eigenvalues as
an ordinary permutator between the corresponding SS. This is actually
what allows to provide the exact spectrum, as we shall see below.

The Fock space $\mathsf{F}$ of the system is ${\mathsf{F}}=
\bigoplus_{N=0}^{2 L} {\mathsf{H}}_N$, where ${\mathsf{H}}_N$ is
the $N$-electron Hilbert space ($\hat{N}=\sum_{i=1}^{L}
\hat{n}_{i \uparrow}+ \hat{n}_{i \downarrow}$). However, due to
the properties of the Hamiltonian, it turns out to be useful to
rearrange ${\mathsf{F}}$ in terms of ${\mathsf{H}}_{N_A}$, {\it
i.e.} the spaces spanned by all the vectors that have a definite
number $N_A$ of sites occupied by states of species $A$
(`$A$-sites' henceforth). Clearly $N_B=L-N_A$. According to the
properties of generalized permutator fulfilled by ${\mathcal{H}}$, the
latter commutes with $\hat{N}_A=\sum_{i=1}^{L} \hat{n}_{i\uparrow}
+\hat{n}_{i\downarrow}-\hat{n}_{i\uparrow}
\hat{n}_{i\downarrow}$, and thus ${\mathsf{H}}_{N_A}$ is preserved
by the dynamics (this would hold in any dimension). In addition,
dealing with an open chain, one can have $3^{N_A}$ possible
sequences ${\mathcal S}$ of $A$-sites for a fixed number $N_A$.
Notice also that, since i) the first term of (\ref{HAM2}) only
permutes $A$ with $B$ and gives zero otherwise, and ii) the
second term merely counts the number of doubly occupied sites,
also the sequence ${\mathcal S}$ is preserved by the dynamics, and
it can be identified with an invariant subspace within
${\mathsf{H}}_{N_A}$. The dimension of each of these $3^{N_A}$
subspaces is ${L \choose N_A}$, accounting for all the possible
actual positions of the $A$-sites along the chain. One can repeat
the above foliation for all the ${\mathsf{H}}_{N_A}$'s ($N_A$ runs
from 0 to $L$) and check that the Fock space is completely
recovered:
\begin{equation}
\sum_{N_A=0}^{L} 3^{N_A}  {L \choose N_A}  \, = \, 4^L \quad,
\end{equation}
so that ${\mathsf{F}}=\bigoplus_{N_A=0}^{L} {\mathsf{H}}_{N_A}$
\\Focusing on a given ${\mathsf{H}}_{N_A}$, one can
characterize each of its basis vector by specifying two
discrete-valued functions: $S(m)$ and $J(m)$ ($m=1, \ldots N_A$).
The former, which is valued 1 (for $|\uparrow\rangle$), 2 (for
$|\downarrow \rangle$) or 3 (for $|\downarrow \uparrow\rangle$)
determines the sequence ${\mathcal S}$ of $A$-sites, and thus the
invariant subspace in which the vector lies; the latter, which is
valued 1 to $L$, determines the actual positions of the $m$-th
$A$-site along the chain. The basis vectors can therefore be
referred to as $|\{S\},\{J\} \, \rangle$, where $`` \{ \, \}''$ is
to remind that $S$ and $J$ are functions.
\\In realizing that the Hamiltonian can be separately diagonalized
within each subspace characterized by a given $A$-sequence
$\mathcal{S}$, it is also crucial to observe that each of such
invariant subspaces can be put in a one-to-one correspondence
with the states of $N_A$ spinless fermion space (or equivalently
with a spin 1/2 model with magnetization $L-N_A$) as follows
\begin{equation}
| \, \{ S \} , \{ J \} \, \rangle \, \, \longleftrightarrow  \,
\, \left( \prod_{m=1}^{N_A} a^{\dagger}_{J(m)} \right) |0 \rangle
\label{trans}
\end{equation}
where $a^{\dagger}$ are the creation operators for a spinless
fermions and $\{ S \}$ the sequence of the subspace.
\\Similarly to what has been done in \cite{AAS} for another extended
Hubbard model, it is also easy to derive the form of an effective
Hamiltonian for the spinless fermion states: indeed, since the
first term in (\ref{HAM2}) reduces to a permutator between SS, it
actually acts on the considered subspace in the same way as a
free Hamiltonian $-t \sum_{\langle i,j\rangle} a^{\dagger}_{i}
a^{}_j$ acts on the spinless problem space. The second term
simply counts the number of species $A$ of kind $|\downarrow
\uparrow \rangle$, namely $N_{\uparrow \downarrow}=\sum_{i=1}^{L}
n_{i\uparrow} n_{i \downarrow} \, \equiv N-N_A$. Therefore the
spectrum in each subspace is given by: \be E(\{n^{A}\};N)= \,
\sum_{k=1}^{L} (-2 t \cos k-U) \, n_{k}^{A} + U N \label{SPE} \ee
where $\{n_{k}^{A} \}$ are quantum numbers valued 0 or 1, $k=\pi
l/(L+1)$ $(l=1, \ldots L)$, and $N$ is the total number of
electrons (which ranges from $N_A$ to $2 N_A$). The eigenvectors
are given by the anti-transformed through (\ref{trans}) of
spinless fermion eigenstates $(\prod_k \, \sum_{i=1}^{L} \sin (k
i) a^{\dagger}_{i} ) |0\rangle$, where the product is over $N_A$
of the $L$ allowed values~of~$k$.
\\When passing from a
subspace of ${\mathsf{H}}_{N_A}$ to another, one finds identical
replica of this spectrum, which amount to having a degeneracy of
the eigenvalues. The degeneracy $g$ corresponds to the different
ways in which one can choose a species $A$ at a given site
provided that $N$ remains unchanged ({\it i.e.} one has the
freedom to change singly occupied $|\uparrow\rangle$ into
$|\downarrow\rangle$ and viceversa); it is therefore easily seen
that
\begin{equation}
g(E(\{n^{A}_{k}\};N))=2^{2 N_A-N} {N_A \choose N-N_A} \quad.
\end{equation}
To conclude this section, we wish to emphasize that the spectrum
(\ref{SPE}) has been derived by means of Sutherland species
technique under {\it open} boundary conditions. In fact the same
model was also studied under {\it periodic} boundary conditions
\cite{AUS}, within the Algebraic Bethe Ansatz approach. However
in the latter case the resulting equations for the quantum
numbers do not allow a straightforward evaluation of the
eigenvalues; indeed the thermodynamics of (\ref{HAM}) had not
been derived yet.
\section{Thermodynamics}
Thanks to the exact spectrum obtained in the previous section, we
can now pass to the study of its thermodynamics, through the exact
calculation of the grand partition function. The language of
Sutherland's species turns out to be very useful to this aim;
indeed, due to the rearrangement of the Fock space described
above, one can write \em \bq {\mathcal Z} & = & Tr \, \left(
e^{-\beta ({\mathcal{H}} -\mu \hat{N})} \right)= \sum_{ \{ n^{A}_k
\} } \, \sum_{N=N_A}^{2 N_A} 2^{2 NA-N} \, {N_A \choose N-N_A} \,
\exp{\left( -\beta \, \left[ \sum_{k=1}^{L} (-2 t \cos k-U)\,
n_{k}^{A} \right] \,
-\beta (U-\mu) N  \right)} = \nonumber \\
 & = & \sum_{ \{ n^{A}_k \} } (2+e^{-\beta(U-\mu)})^{N_A} \exp{ \left(
\sum_{k=1}^{L} \left[ \beta (2 t \cos k+\mu) \right] \, n_{k}^{A}
\right) }
 \, = \, \prod_{k=1}^L  \left( 1+ \exp \left[ \beta \,
 (2 t \cos k+\mu+\nu(U,\beta,\mu) \, ) \right]  \right)
\label{zeta} \eq \bm \noindent where we have defined $\nu(U,\beta,
\mu) =\ln (2+e^{-\beta(U-\mu)})/ \beta$, \, \, $\beta=1/(k_B T)$
being the inverse temperature and $\mu$ the chemical potential as
usual.
\\The grand-potential is easily obtained as $\omega=\omega(\beta;U;\mu)=-
\lim_{L \rightarrow \infty} \beta^{-1} (\ln {\mathcal Z} \, /L)$. After
introducing $\mu_{eff}=\mu+\nu$, $\omega$ reads
\end{multicols}
\be \omega(\beta;U;\mu) \, = \, -  \, \frac{1}{\pi \, \beta}
\int_{0}^{\pi} \, dk  \, \, \ln \left( 1+ \, \exp \left[ \beta \,
(2 t \cos k + \mu_{eff}(U,\beta,\mu)\, ) \right] \, \right)
\label{omega} \ee \bm \noindent Remarkably, the grand potential is
formally similar to that of a tight-binding model with an {\it
effective} chemical potential $\mu_{eff}$. We stress that
$\mu_{eff}(U,\beta,\mu)$ depends on the on-site Coulomb repulsion, the
temperature and the chemical potential, in a highly
non-linear way. This yields peculiar features of the model, as we
shall show in the following.
\\In deriving the thermodynamics of the system, it is
customary to eliminate $\mu$ in favor of the filling $\rho$; the
latter can be computed as $\rho=-\partial \omega / \partial \mu$,
and the result turns out to be of the following form
\begin{equation}
\rho(U,\beta,\mu) \, = \, (1+C(U,\beta,\mu)) \cdot
\rho_A(\beta,\mu_{eff} (U,\beta,\mu)) \label{fill}
\end{equation}
where:
\begin{equation}
C(U,\beta,\mu)=\frac{\exp{(-\beta (U-\mu))}}{2+\exp{(-\beta
(U-\mu))}}
\end{equation}
and
\begin{equation}
\rho_A(\beta,\mu_{eff})= \frac{1}{\pi} \int_{0}^{\pi} \,
\frac{dk} { 1+ \, \exp \left[ \beta(-2 t \cos k-\mu_{eff}) \right]
} \label{rho_A}
\end{equation}
Notice that differentiating $\omega$ with respect to $\mu_{eff}$
instead of to $\mu$ would yield only the right factor $\rho_A$ of
(\ref{fill}); the non-linearity of $\nu$ as a function of $\mu$
results in the appearance of $C$ in the left factor; this causes
the relation $\mu=\mu(\rho;T;U)$ implicitly defined by
(\ref{fill}) to be very different from that of a tight binding
model, as we shall explicitly show in next section.
\\The two factors in (\ref{fill}) deserve some comment; $\rho_A$ is
nothing but the density of $A$-sites along
the chain, defined as $\rho_A=\lim_{L \rightarrow +\infty}
\langle \hat{N}_A \rangle /L$; the functional dependence of
$\rho_A$ on $\beta$ and $\mu_{eff}$ is that of a spinless
tight-binding model. The left factor provides information,
through the function $C$, about the kind of occupancy of the
sites of the chain; indeed when $C \sim 0$ most of the occupied
sites are singly ($s$) occupied, whereas if $C \sim 1$ most of
the occupied sites are doubly ($d$) occupied; intermediate values
indicate the percentage of $d$ with respect to $s$
sites.
\\To conclude this section we wish to comment about the energy
(per site) of the system; the latter is obtained by
${\mathcal{E}}=-\lim_{L \rightarrow \infty}
\partial(\ln {\mathcal{Z}} \, / L)/ \partial \beta +\mu \rho$ and
reads \em
\begin{equation}
{\mathcal{E}}(U,\beta,\mu)=\frac{1}{\pi} \int_{0}^{\pi}  \, dk \,
\, \frac{-2 t \cos k -U}{1+ \, \exp \left[ \beta(-2 t \cos
k-\mu_{eff}(U,\beta,\mu)) \right] } \, \, + \, U \rho \quad .
\label{energy}
\end{equation}
\bm Eq.(\ref{energy}) naturally allows to identify in
$\mathcal{E}$ a kinetic energy $\mathcal{K}$ and a potential
energy $\mathcal{P}$. The former is defined as the weighted
integral of $-2 t \cos{k}$ and the latter as the weighted
integral of $-U$, which actually gives $-U\rho_A$, according to
(\ref{rho_A}). In fact the actual potential energy would also
contain the last term $U \rho$ of (\ref{energy}); however, since
this is merely a constant with respect to temperature, we prefer
not to include it in the definition of $\mathcal{P}$, so that the
latter describes the only temperature dependent part of the
potential term $U \hat{n}_{i \uparrow} \hat{n}_{i \downarrow}$.
Notice that with this choice the potential energy is attractive
for positive $U$. Notice also that, although $\mathcal{K}$ and
$\mathcal{P}$ are clearly related to the hopping terms and to the
on-site Coulomb repulsion respectively, they are not mutually
independent: indeed $\mathcal{K}$ depends not only on $t$ but
also $U$ and viceversa for $\mathcal{P}$. We shall come back to
this point in discussing the specific heat in section V.
\subsection{Chemical potential}
\begin{figure}
\epsfig{file=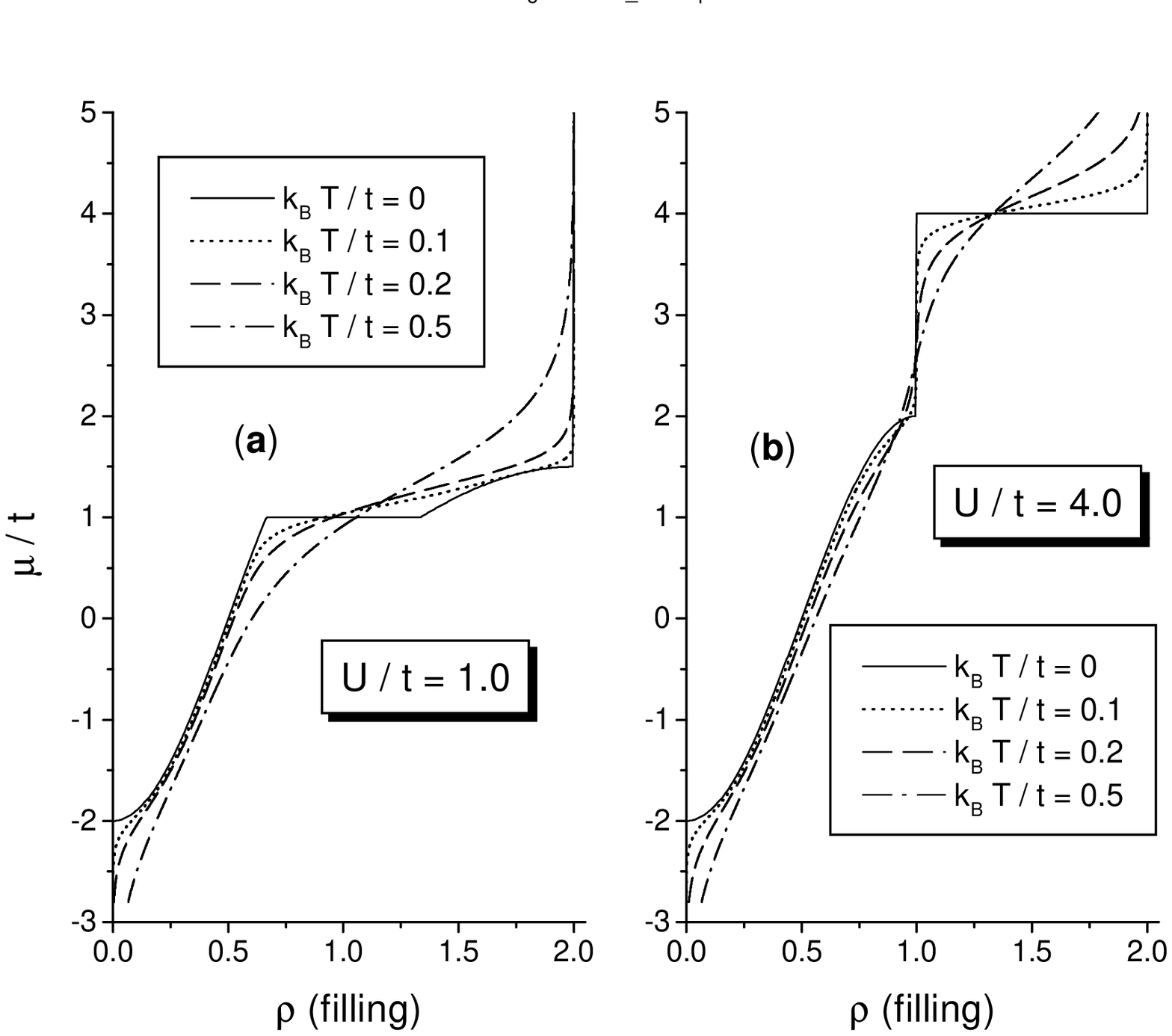,width=7.5cm,height=5.8cm,clip=}
\caption{The relation between $\rho$ and $\mu$ for different
temperatures, at $U/t=1$ (a), and $U/t=4$ (b). For $T=0$, the
curve in fig.(a) shows a plateaux, related to 'phase coexistence'
in mixed region II of fig. \ref{ter_fig1}; whereas the curve in
fig.(b) exhibits a jump in $\mu$, due to the opening of charge
gap at half-filling.} \label{ter_mu1}
\end{figure}

The chemical potential $\mu(\rho;T;U)$ of our model is shown in
fig.\ref{ter_mu1} at $U=t$ (a) and $U=4 t$ (b), for
different values of the temperature.
\\Focussing first on the solid curves, representing the case $T=0$, one
can realize that even in the ground state the relation between $\mu$
and $\rho$ is quite different from that of a spinless tight-binding
model, which would read $\mu(\rho;T=0)=-2 t \cos( \pi \rho)$.
\\In particular, in fig.(a) we notice that a `plateaux' appears, in
correspondence with region II of the ground state phase diagram (see
fig.\ref{ter_fig1}). Interestingly, such shape reminds that of a
coexistence region, connecting the phase of single carriers (region
III-a) to that of pair-carriers (region I); this would imply that, as
the filling is
increased, the 1D lattice starts exhibiting macroscopic regions
made of only single carriers separated by other macroscopic
regions where only pairs are present. In fact, eigenstates with
such features are certainly present; however, they are degenerate
with other eigenstates, in which single and pair carriers
alternate with no macroscopic order. This is basically due to the
degeneracy of $A$-sequences in such region.
\\In fig.(b) a vertical jump is instead present at half-filling,
as an hallmark of the opening of the charge-gap. The flat part of
the solid curve for $\rho>1$ just coincides with the atomic limit
behavior (region III-b of fig.\ref{ter_fig1}).
\\Considering now the curves at finite temperature of fig.\ref{ter_mu1},
one can observe how the edges present at $T=0$ smoothen as soon as
$T>0$. A remarkable feature is the presence in fig.(b) of a nearly
universal point ($\rho^{*}=4/3,\mu^{*}=U$), where all the curves
of sufficiently low temperatures basically intersect. Such kind of
points are in general determined through the conditions $\partial
\mu/\partial T=0$ and $\partial^2 \mu/\partial T^2 \,=0$. It is
in fact possible to calculate that for any $U> 2t$ and $\rho>1$
(region III-b of fig.\ref{ter_fig1}) the low-temperature behavior
of $\mu$ is given by
\begin{equation}
\mu \simeq U+ k_B T \ln \left( \frac{2 (\rho-1)}{2-\rho} \right) +
{\mathcal{O}}(e^{-\frac{U-2t}{k_B T}}) \quad, \label{mu-low-T1}
\end{equation}
whence the above conditions are both fulfilled up to
exponentially small terms in $k_BT/t$. \\We shall also see in
subsection D that nearly-universal crossing points are exhibited
by other observables of the model, such as the specific heat.
Eq.(\ref{mu-low-T1}) also points out that in our model a {\it
linear} low-temperature behavior is possible, differently from
the tight-binding model, where only even powers in $T$ are
allowed in the Sommerfeld expansion. In general, in our model,
different behaviors of $\mu$ arise according to the values of $U$
and $\rho$. For instance, for $U$ and $\rho$ belonging to the
mixed region II of fig.\ref{ter_fig1}, the chemical potential at
low-temperature has again a linear term,
\begin{equation}
\mu \simeq U+k_B T \ln \left( \frac{2 (\rho-\bar{\rho})}{2
\bar{\rho}-\rho} \right) + {\mathcal{O}}((k_B T/t)^2) \quad,
\label{mu-low-T}
\end{equation}
but with a coefficient which depends on $U$, since
$\bar{\rho}=\bar{\rho}(U)=\pi^{-1} \cos^{-1}(-U/2t)$. \\In
contrast, when the charge gap $\Delta_c=U-2t$ opens ({\it i.e.}
at $\rho=1$ and $U>2t $), $\mu$ acquires a highly non-linear form
\begin{equation}
\mu  \simeq 2t+ \frac{\Delta_c}{2}+\frac{k_B T}{4 t} \ln(\frac{k_B
T}{4 \pi t}) \quad,
\end{equation}
indicating that the behavior is definitely different to that of
an intrinsic semiconductor.
\begin{figure}
\epsfig{file=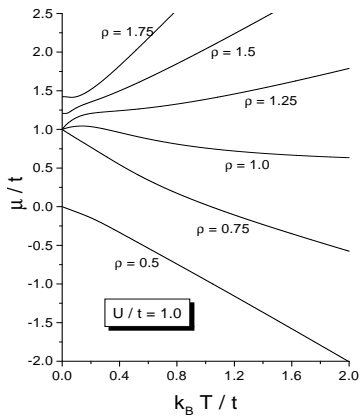,width=5cm,height=6cm,clip=} \caption{The
behavior of $\mu$ as a function of temperature, for fixed $U/t=1$
and different values of filling. The asymmetry with respect to
the half-filled case $\rho=1$ is ascribed to the lack of particle-hole
invariance of the model.} \label{ter_mu2}
\end{figure}
\noindent In fig.\ref{ter_mu2} we explicitly examine the behavior
of $\mu$ as a function of temperature, for a fixed value of
on-site Coulomb repulsion ($U/t=1$) and for different fillings. A
main difference has to be emphasized with respect to the case of a
tight-binding model: in the latter the curves of $\mu$ are
specular for filling values that are symmetric with respect to
half filling ({\it i.e.} $\mu \rightarrow -\mu$ for $\rho
\rightarrow 2-\rho$), whereas this is not the case in our model,
due to the fact that it is not particle-hole invariant.
\subsection{Compressibility}
\begin{figure}
\epsfig{file=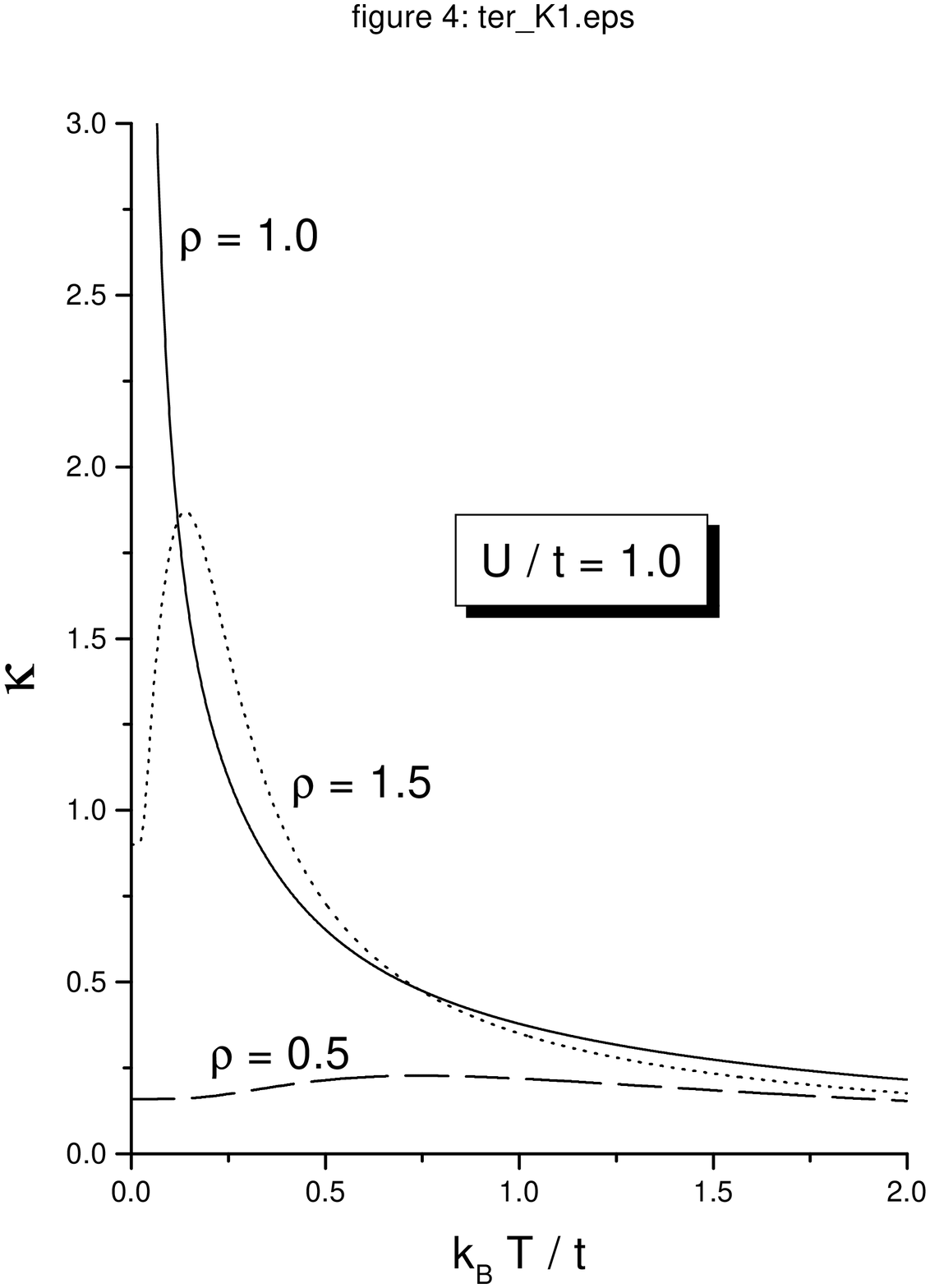,width=7cm,height=8cm,clip=} \caption{The
compressibility as a function of temperature, at fixed on-site
Coulomb repulsion $U/t=1.0$ and for different filling values. $\kappa$
diverges as $T\rightarrow 0$ for values of $U/t$ and $\rho$ belonging to
mixed region II of fig. \ref{ter_fig1}.}
\label{ter_K1}
\end{figure}

\begin{figure}
\epsfig{file=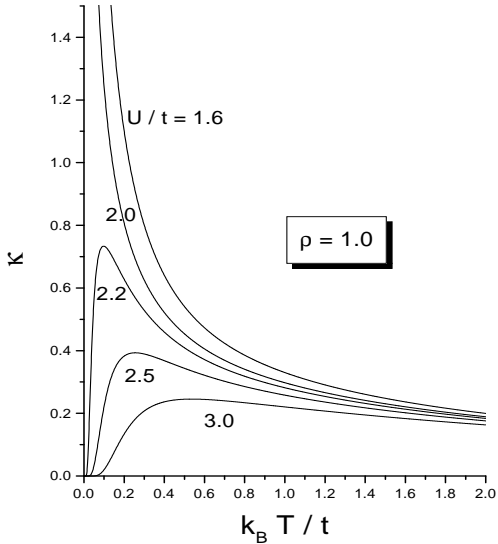,width=7cm,height=8cm,clip=} \caption{The
compressibility as a function of temperature, at fixed filling
$\rho=1$ and for several values of on-site Coulomb repulsion. As the
charge-gap opens ($U>2 t$) $\kappa$ acquires an exponential
low-temperature behavior.}
\label{ter_K2}
\end{figure}

The compressibility $\kappa=\partial \rho / \partial \mu$ can be
easily evaluated through eq.(\ref{fill}). In fig.\ref{ter_K1} we
have plotted $\kappa$ as a function of the temperature, for a
fixed value of $U$ (namely $U/t=1.0$) and for different fillings.
One can observe the change in the low-temperature behavior when
tuning the filling: at $\rho=0.5$ the behavior is regular, while
at half-filling $\kappa$ undergoes a singularity for $T
\rightarrow 0$; eventually ($\rho=1.5$) its behavior is again
regular. The reason for the low-temperature singularity at
$\rho=1$ is that in the ground state the point ($U/t=1; \rho=1$)
is situated in region II (see fig.\ref{ter_fig1}), {\it i.e.} in
the region where the chemical potential exhibits the plateaux, as
shown in fig.\ref{ter_mu1}; such singularity is indeed present
for all values of $U$ and $\rho$ that belong to that region of the
ground state. The divergence of $\kappa$ can be proved to be of
the type $\propto T^{-1}$. \\In contrast the behavior for $T
\rightarrow 0$ at $\rho=0.5$ and $\rho=1.5$ is regular since such
filling values belong to regions III-a and I respectively.
In fig.\ref{ter_K2} we have examined in detail the case of
half-filling, plotting $\kappa$ as a function of $T$ for different
values of $U$; one can explicitly observe how $U=2t$ is the
critical value separating the divergent behavior for $U<2t$ from
the regular one for $U>2t$. Indeed, as soon as $U>2t$, the
divergence becomes a pronounced peak in $\kappa$; the temperature
$T^{*}$ at which the peak occurs increases with increasing $U$,
similarly to what happens in the ordinary
Hubbard model, according to the results \cite{JKS}. Notice that
in contrast no singular behavior is expected at moderate $U's$ in
the ordinary Hubbard model at half-filling, since in that case the
system is insulating for any positive $U$.
\subsection{Local magnetic moment}
\begin{figure}
\epsfig{file=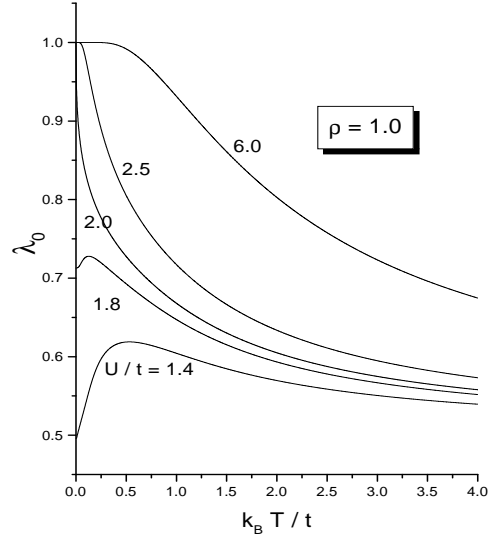,width=7cm,height=8cm,clip=}
\caption{Local magnetic moment as a function of the temperature,
at half filling for several values of $U/t$. Notice how the
low-temperature behavior abruptly changes across the
metal-insulator transition point. The figure indirectly provides
also the behavior of $\rho_A$, since $\lambda_0=2 \rho_A -\rho$.}
\label{ter_LMM1}
\end{figure}

The local magnetic moment was first introduced in \cite{PINCUS}
and is defined as
\begin{equation}
\lambda_0=\lim_{L\rightarrow \infty} \langle \frac{1}{L} \sum_{j}
(\hat{n}_{j \uparrow}-\hat{n}_{j \downarrow})^2 \rangle \quad .
\label{LMM}
\end{equation}
It characterizes the magnitude of spin at each site, {\it i.e.}
the degree of localization of electrons. In terms of density of
$A$-site, $\lambda_0$ can be easily rewritten as $\lambda_0 =
 \rho-2
\rho_{\uparrow \downarrow}=2 \rho_A-\rho$, where $\rho_A$ can be
computed from eq.(\ref{rho_A}). \label{sub-LMM} In
fig.\ref{ter_LMM1} we have reported the local magnetic moment at
half filling for different values of the on-site Coulomb
repulsion. One can observe that the behavior of $\lambda_0$, even
within a relatively small range of values of $U$, is quite rich.
In order to describe it, we first consider the case of small
values of $U$ (namely $U=1.4 t$ in the figure); we recall that in
the ground state such value corresponds to the mixed region II
(see fig. \ref{ter_fig1} at $\rho=1$), meaning that hopping paired
electrons are present at $T=0$; as the temperature is turned on,
$\lambda_0$ first increases with $T$ (indicating that the pairs
are broken in favor of single carriers); however, after reaching a
maximum at a temperature $T^{*}$, $\lambda_0$ starts decreasing
for higher $T$'s, denoting that pairs are now reformed by higher
thermal excitations. According to the above observations, it easy
to realize that the temperature $T^{*}$ decreases with increasing
$U$; in fact the maximum disappears for $U \simeq 1.85$, so that
$\lambda_0$ becomes a definitely decreasing function of the
temperature. At $U=2t$, $\lambda_0$ reaches at $T=0$ the
saturation value 1 (all singly occupied sites), with an infinite
derivative with respect to the temperature. Passing through
$U=2t$, an abrupt change in the low-temperature slope occurs: the
curves of $\lambda_0$ suddenly flattens as soon as $U>2t$. This
reflects the metal-insulator transition occurring in the ground
state; indeed the opening of the charge gap causes the formation
of pairs to be highly unfavored at low $T$'s.
\subsection{Specific heat}
\begin{figure}
\epsfig{file=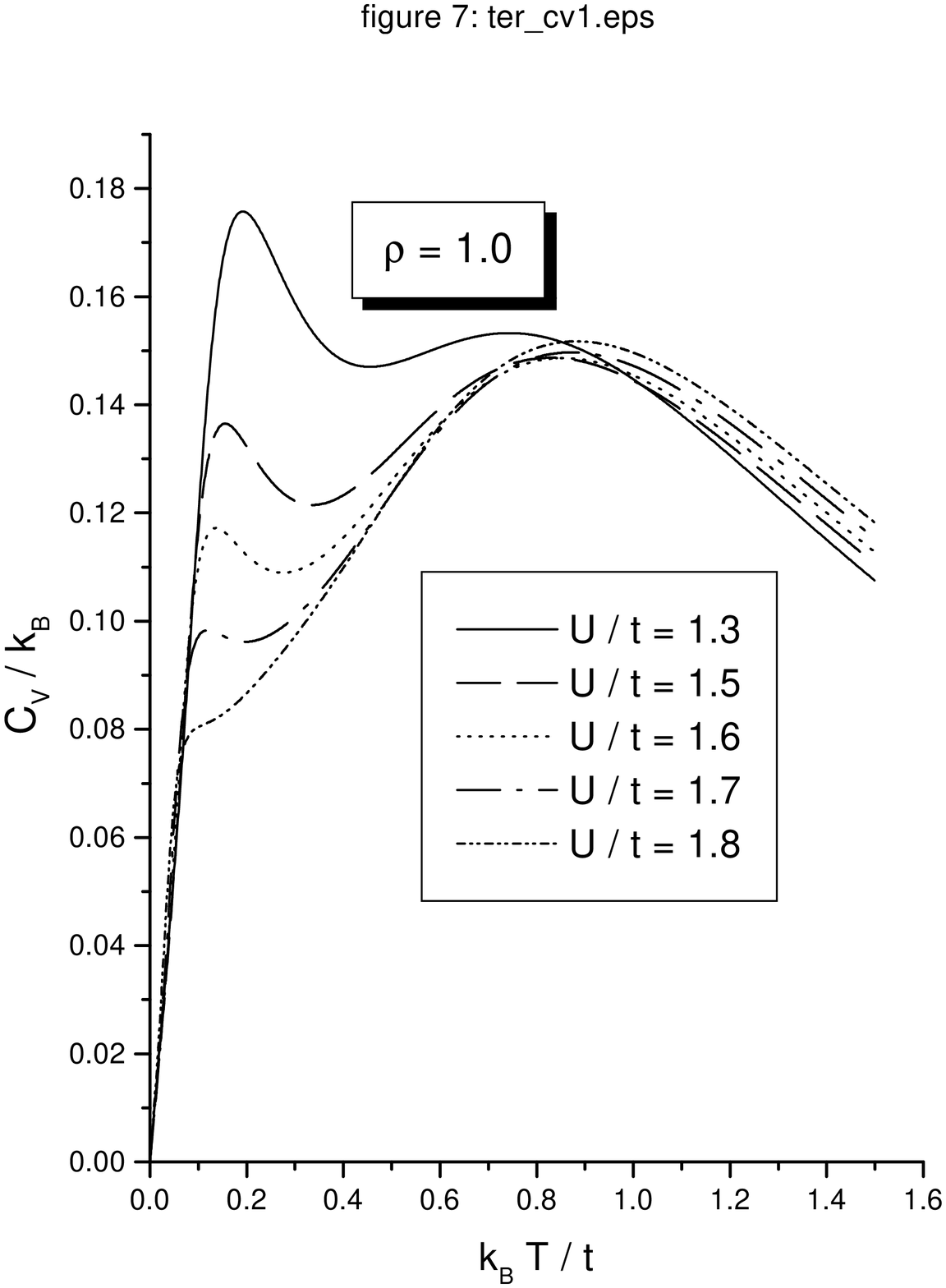,width=7cm,height=8cm,clip=}
\caption{Specific heat as a function of $T$ at half filling, for
different values of $U/t$ below the metal-insulator transition
value: a two-peak structure is present, as well as a nearly
universal crossing point.The low-temperature behavior is linear.}
\label{ter_cv1}
\end{figure}

\begin{figure}
\epsfig{file=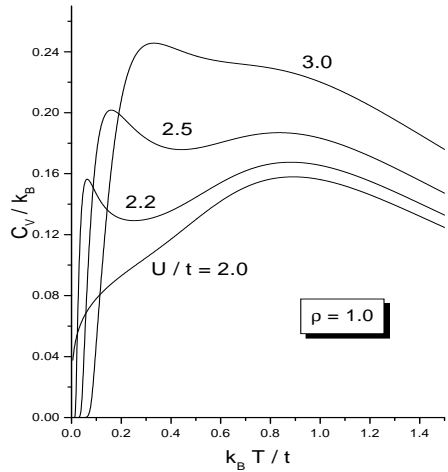,width=6.5cm,height=7cm,clip=}
\caption{Specific heat as a function of $T$ at half filling and
different values of $U/t$, just above the metal-insulator
transition point: the two-peaks structure definitely disappears
for $U\gtrsim 3 t$. The low-temperature behavior is exponential
for $U>2 t$.} \label{ter_cv2}
\end{figure}

\begin{figure}
\epsfig{file=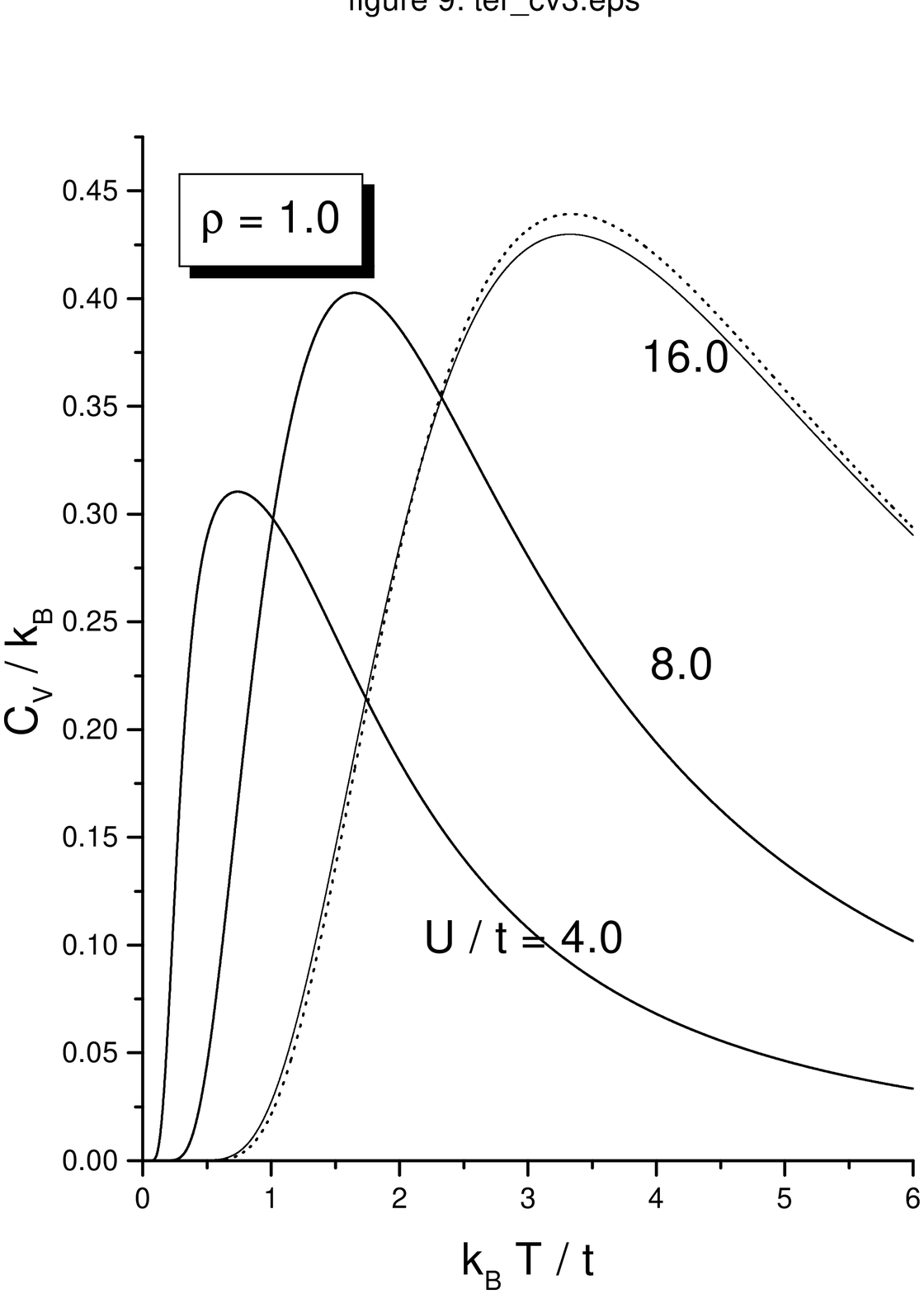,width=6.5cm,height=7cm,clip=}
\caption{Specific heat as a function of $T$ for half-filling and
$4t \le U \le 16 t$. In the strong coupling regime, the two peaks
structure disappears: the remaining peak is well described from
the atomic limit model (dotted curve). A similar behavior is
obtained also at any $\rho> 1$.} \label{ter_cv3}
\end{figure}

\begin{figure}
\epsfig{file=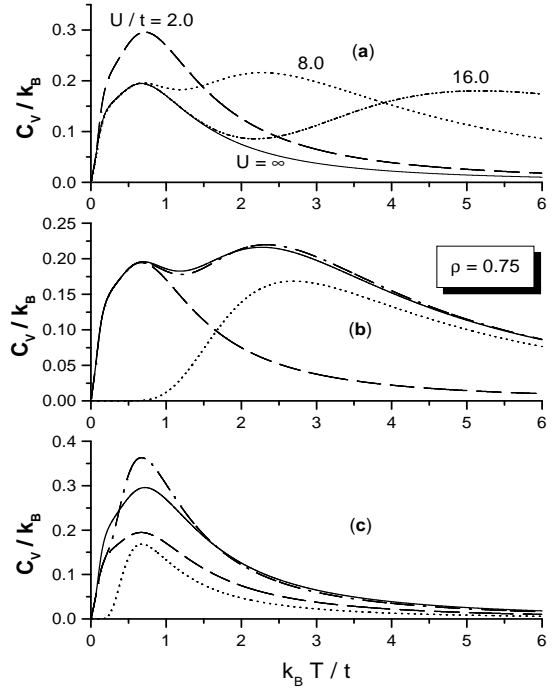,width=8cm,height=10cm,clip=}
\caption{The specific heat as a function of temperature for
$\rho=0.75$. In fig.(a) $C_V$ is plotted for different values of
$U/t$; (b) at strong coupling ($U/t=8$) the specific heat of the
model (solid line) is well reproduced by the sum (dot-dashed) of
the specific heat of $U=\infty$ model (dashed) and of the atomic
limit (dotted); (c) this is not the case at moderate coupling,
where the energy scales of the two models become comparable
($U/t=2$).} \label{ter_cv4}
\end{figure}

In this subsection we present our results on the specific heat of
model (\ref{HAM}) which can be computed through
\begin{equation}
C_V= \frac{d \mathcal{E}}{d T} =  -k_B \beta^2 \left(
\frac{\partial \mathcal{E}}{\partial \beta} - \frac{\partial
{\mathcal{E}}}{\partial \mu} \frac{\partial \rho}{\partial \beta}
/ \frac{\partial \rho}{\partial \mu} \right) \label{cv} ,
\end{equation}
where the energy ${\mathcal{E}}$ is given by (\ref{energy}). Below
we study the temperature dependence of $C_V$ when varying the
physical parameters $U$ and $\rho$. The exact calculation shows
that in our model a two-peak structure is definitely present not
only in the strong coupling regime, but also at moderate $U$'s.
\\We start considering the case of half-filling ($\rho=1$).
The two peaks appear first for $1.3 \lesssim U/t \lesssim 1.8$  (see
fig.\ref{ter_cv1}); in this range of $U$'s, $C_V$ also exhibits a
nearly universal crossing point at $k_B T \sim 0.85 t$; we shall
comment on such feature at the end of this subsection. The
peaks eventually merge into one for $U/t \sim 1.85$. However, as
soon as $U> 2t$ (see fig.\ref{ter_cv2}), a new well pronounced
low-temperature peak appears. The recovered double-peak structure
is present up to $U \sim 3t$, where finally only one peak survives.
\\By comparing figs.\ref{ter_cv1} and \ref{ter_cv2}, one can notice that
the metal-insulator transition point $U=2t$ is also the hallmark
of a crossover in the low-temperature behavior of $C_V$. In
particular, the calculation shows that for $U<2 t$ the latter is
linear,
\begin{equation}
C_V \simeq \frac{k_B}{2 \pi \sqrt{1-(\frac{U}{2 t})^2} } \left(
\frac{\pi^2}{3}+\ln^2 \frac{4 (1-\bar{\rho}) \bar{\rho}}{(2
\bar{\rho}-1)^2} \right) \cdot \frac{k_B T}{t} \, \, \quad,
\end{equation}
where $\bar{\rho}$ is defined as in eq.(\ref{mu-low-T}). In
contrast, for $U>2 t$, $C_V$ exhibits an exponential-like behavior
given by
\begin{equation}
\displaystyle C_V \simeq \frac{k_B}{(4 \pi)^{1/4}}
(\frac{\Delta_c}{2t})^2 \left(\frac{k_B T}{t}\right)^{7/4}
\exp(-\frac{\Delta_c}{2 k_B T}) \quad,
\end{equation}
where $\Delta_c=U-2t$ is the charge gap.
To conclude the study at half-filling we have examined the case
of large $U/t$ (see fig.\ref{ter_cv3}). The result shows that
only one peak is present, at a temperature which increases almost
linearly with $U$ ($k_B T\sim 0.21 \, U$). This result can be
understood considering that at large $U/t$ the spectrum
(\ref{SPE}) exhibits two different energy scales: i) a low-energy
scale ($\sim t$), which describes fluctuations in the $A$-band,
whose effective filling is given by the value of $\rho_A$; ii) a
high-energy scale (of the order of $U$) involving the formation
of on-site pairs, favoring the decrease of the number of
$A$-sites. The former channel is actually active only for
$\rho<1$, since at half-filling the $A$-band becomes completely
filled: indeed in this case we have $\rho_A\simeq 1$ for $k_B T
\sim t$, as can be deduced from fig. \ref{ter_LMM1} of the local
magnetic moment at large $U/t$.
\\Only the high-energy channel is thus active, and its contribution is
well described by the atomic-limit model ({\it i.e.} $t=Y=0$),
shown by the dotted curve in fig.\ref{ter_cv3}. The slight
deviations are due to the fact that, as pairs are formed from
singly occupied sites via thermal fluctuations, the number of
effective species A decreases, and the formed A-holes can produce
(relatively small) fluctuations with~$T$. However, the larger is
$U/t$, the better is the agreement with the specific heat of the
atomic limit.
\\We also wish to emphasize that the behavior is different
from that of the ordinary Hubbard model, where two peaks
appears at low temperatures in the strong coupling limit at
half filling. In fact, although in the Hubbard model the lower
Hubbard band is filled, spin excitations of low-energy ($\sim
J=4t^2/U$) are active. These kind of excitations are instead
absent in our model; we shall comment in more detail in section V
about this point.
In fig.\ref{ter_cv4} we investigate the specific heat for filling
values below half-filling, namely $\rho=.75$.
\\As fig.\ref{ter_cv4}-(a) shows, a double peak structure of $C_V$
appears; however, two important differences have to be emphasized
with respect to the case of half-filling: in the first instance,
here the double-peak structure arises and becomes more evident
for {\it large} values of $U$'s, whereas at half filling it is
present at {\it moderate} $U$'s; secondly, the temperatures of
the two peaks are quite higher than the corresponding ones of the
half filled case. In particular the position of the
low-temperature peak is practically independent of $U$, whereas
the high-temperature one strongly depends on it, similarly to
what happens for the only peak present at half-filling in the
strong coupling regime (fig.\ref{ter_cv3}).
\\The two peaks of fig.(a) have to be related to the
two energy scales emerging in the spectrum when $U \gg t$, as
discussed above; in particular, the low-temperature one is
attributed to the fluctuations of the $A$-band, which is now
partially filled, unlike for half-filling. We recall that in this
range of the parameters $U$ and $\rho$, the ground state of the
model is that of the $U=\infty$ model (region III-a of
fig.\ref{ter_fig1}); since the formation of pairs is strongly
inhibited for high $U$'s, the physics of low-energy excitations
is fairly captured by that of the $U=\infty$ model at finite
temperature, as shown by the solid curve in fig.(a). In fig.(b)
the case $U/t=8$ is examined in detail; in this case the sum of
the specific heats of $U=\infty$ model and atomic model
practically recovers the actual $C_V$ of our model. Such
agreement improves with increasing $U$, whereas at moderate
values of $U$ the argument of energy scale separation does not
hold: indeed the high temperature peak merge into the
low-temperature one for $U \sim 2t$, and $C_V$ is no more given
as the sum of $U=\infty$ and atomic limits (see fig. (c)).
We have also considered the case of filling values greater than
one. In the strong coupling regime the ground state has the
$A$-band completely filled, the sites of the chain being all
occupied (either singly or doubly, as shown in region III-b of
fig.\ref{ter_fig1}); the low-energy scale is thus frozen, just
like in the case of half-filling. This yields the specific heat
behavior be actually described by that of the atomic limit,
similarly to fig.\ref{ter_cv3}. The temperature of the peak grows
linearly with $U$ ($k_B T \simeq c(\rho) U$), the
coefficient $c$ being an increasing function of the filling $\rho$.\\

\begin{figure}
\epsfig{file=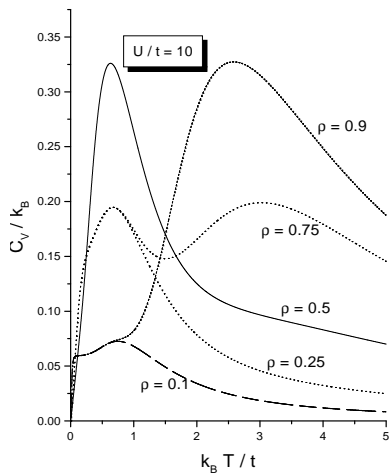,width=5.5cm,height=7cm,clip=}
\caption{The specific heat as a function of temperature at strong
coupling and $\rho<1$. The low-temperature behavior is the same
for values of $\rho$ symmetric with respect to $\rho=0.5$: indeed
in this case low-energy excitations are well described by the
$U=\infty$ Hubbard model, which is particle-hole invariant around
quarter filling. Differences instead emerge at high temperatures.}
\label{ter_cv5}
\end{figure}

\begin{figure}
\epsfig{file=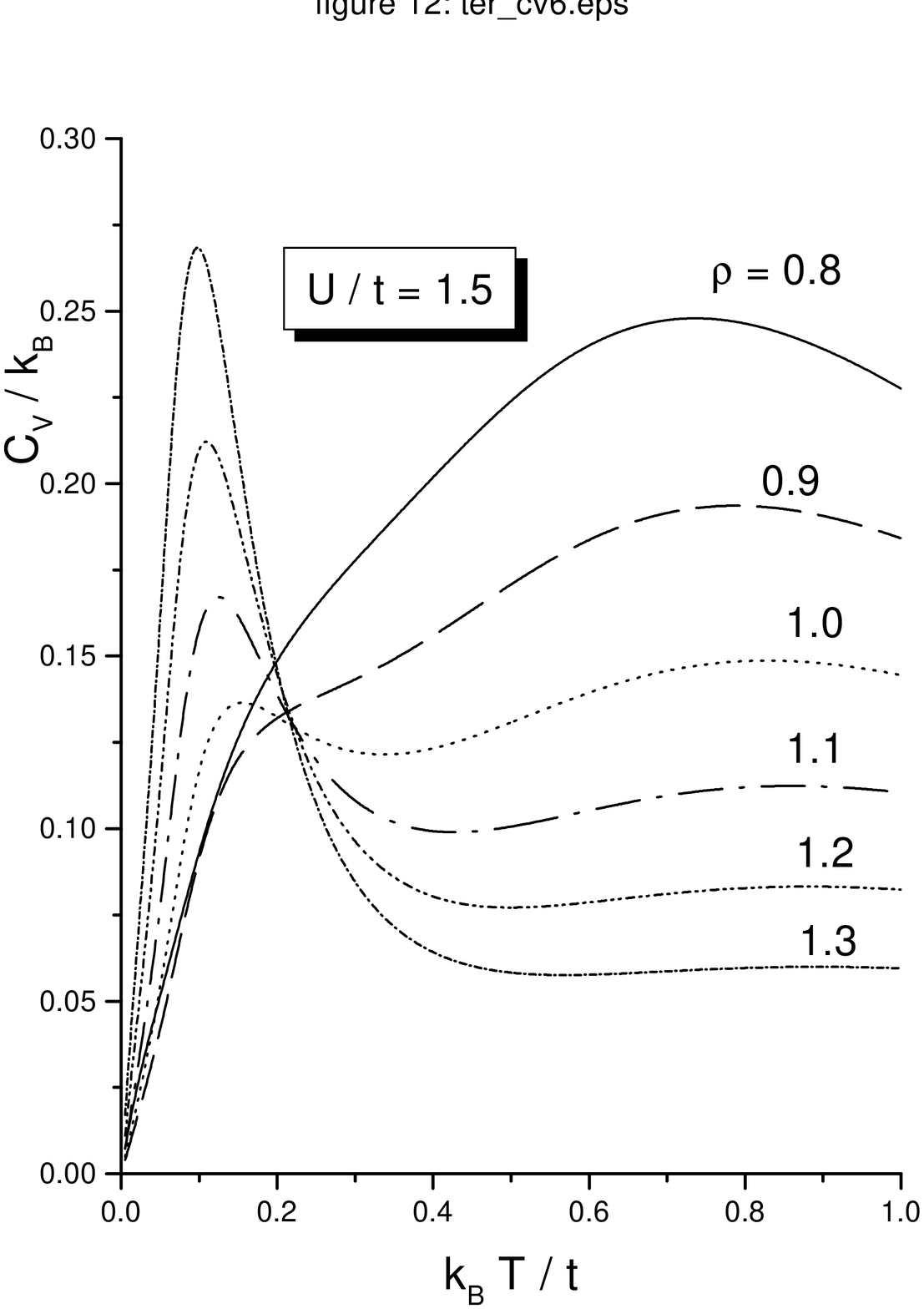,width=5.5cm,height=7cm,clip=}
\caption{The specific heat as a function of temperature in the
moderate $U$'s regime. A nearly universal crossing point with
varying $\rho$ at fixed $U$ is observed for values of $\rho$ in
the range $0.9\lesssim \rho\lesssim 1.3$.} \label{ter_cv6}
\end{figure}

Figs.\ref{ter_cv5} and \ref{ter_cv6} examine the filling
dependence of the specific heat at fixed coupling values. More
precisely, fig. \ref{ter_cv5} reports the results obtained in the
strong coupling case. As anticipated above, in this case the low
temperature peak is perfectly recovered from the $U=\infty$
model; notice that, since the latter is particle-hole symmetric
around quarter filling ($\rho=0.5$) the low-temperature behavior
of curves related to filling values that are symmetric with
respect to $\rho=0.5$ is basically identical. In contrast, the
higher temperature peak does not exhibit such symmetry, being
related to the atomic limit of the Hubbard model, which is no
more particle-hole symmetric around quarter filling.
\\Fig.\ref{ter_cv6} is concerned with the behavior at moderate
$U$'s (namely $U/t=1.5$) as a function of $\rho$; the remarkable
feature is the appearance of a nearly universal crossing point at
low temperature ($k_B T\sim 0.2 \, t$), for a finite range of
filling values ($1.0 \lesssim \rho \lesssim 1.3$). Similarly, a
nearly universal crossing point also occurs at fixed filling for
varying $U$, as fig.\ref{ter_cv1} shows. The latter type of
behavior is also exhibited by the ordinary half-filled Hubbard
model\cite{JKS,GEB,DUFFY,GEO}; however, to the authors'
knowledge, theoretical investigations were mostly limited to the
case of fixed filling and varying $U/t$. In contrast, here we have
explored the case of varying $\rho$ as well; this is interesting
in view of comparison with experimental results, where $U/t$ can
be roughly interpreted as the inverse pressure, and
$\delta=|1-\rho|$ as the doping. In fact, this type of universal
behavior has been observed in many heavy-fermion compounds, such
as cerium ones, both at fixed doping with varying
pressure\cite{HEAVY1}, and at fixed pressure with varying
doping\cite{HEAVY2}. Let us notice that, for the ordinary Hubbard
model, the presence of the nearly universal point in $U$ has been
explained in \cite{VOL}, as a consequence of the fact that the
entropy $S$ at high temperatures does not depend on $U$, in that
case. For our model, $S$ at high temperatures is also independent
of $U$; however it turns out that it {\it does} depend on $\rho$.
Hence we expect that the argument in\cite{VOL} cannot be applied
to explain the nearly universal crossing point in $\rho$ shown in
fig.\ref{ter_cv6}.
\begin{figure}
\epsfig{file=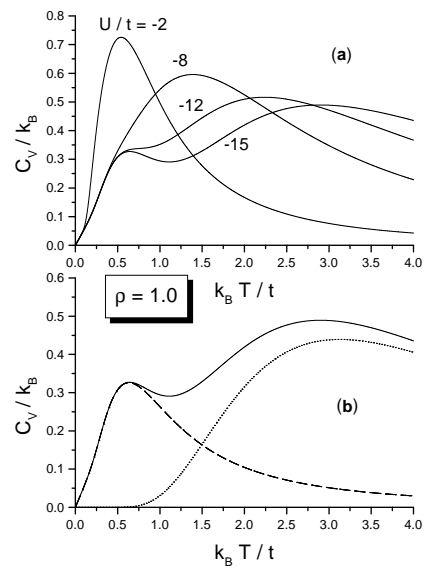,width=6cm,height=8cm,clip=} \caption{The
specific heat as a function of temperature at half filling for
negative values of $U$. Fig.(a): the double peak emerges as $|U|$
is increased. Fig.(b): in the strong coupling regime
($U/t=-15.0$) $C_V$ is fairly reproduced by the sum of the XX0
contribution (dashed line) and the atomic limit contribution
(dotted line).} \label{ter_cv7}
\end{figure}
Finally, the specific heat $C_V$ is investigated in
fig.\ref{ter_cv7} also for negative values of the Coulomb
interaction, at half-filling. The behavior is quite different
with respect to the positive $U$ case for moderate and
intermediate $U$ values, since no double peak is present. \\In
contrast, such structure emerges at higher coupling values; also
in this case two separate energy scales emerge. However, the low
temperature peak is now reproduced by that of the $XX0$ model
($t=0$), whose ground state actually coincides with that of our
model, for these values of $U$ and $\rho$ (see region I in
fig.\ref{ter_fig1}). The high-temperature peak is still due to
the negative-$U$ atomic limit ($t=Y=0$). In figure (b) it is
clearly shown how, in the strong coupling case, the simple sum of
the specific heats of $XX0$ and atomic limit perfectly reproduces
the result for our model; this is not the case by at lower $U$
values. \label{sub-cv}
\section{Discussion}
As outlined in the previous section, our results show that the
specific heat exhibits a two-peak structure for different values
of on-site Coulomb repulsion $U$ and filling $\rho$. In the
present section we wish to discuss the origin of the two peaks,
since in last years much effort has been made to clarify a
similar behavior occurring in the ordinary Hubbard model. As
mentioned in the introduction, in the latter model the two peaks
are usually explained in terms of `spin' and `charge' excitations.
\\The above argument cannot be applied here, since our model involves
only `charge' degrees of freedom: in fact, from the {\it formal}
point of view of quantum numbers $n^{A}_k$, the excitation
processes in the spectrum (\ref{SPE}) have the typical feature of
`charge' excitations (in the sense of $A$-species). It is however
worth emphasizing that, just like for the ordinary Hubbard model,
the nomenclature based on quantum numbers does not strictly
correspond to its {\it physical} meaning. In our case, the
`charge' degrees of freedom of $A$-species actually carry both
{\it charge} and {\it spin} density fluctuations: the break-up of
a localized pair into two single carriers indeed leads to a
redistribution of the charge density as well as to the formation
of a triplet replacing a singlet state. \\

\begin{figure}
\epsfig{file=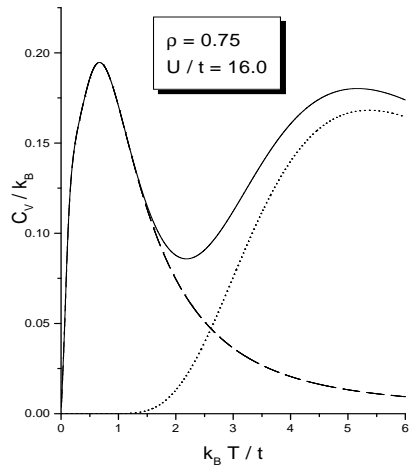,width=6cm,height=7cm,clip=} \caption{The
kinetic (dashed) and potential (dotted) contributions to the
specific heat (solid), at strong coupling ($U/t=16$), for
$\rho=0.75$. The low-temperature peak is basically due to
$\mathcal{K}^{\prime}$, while the high-temperature peak stems
from $\mathcal{P}^{\prime}$. In this regime ($U \gg t$),
$\mathcal{K}^{\prime}$ is also well described by the specific
heat of $U=\infty$ Hubbard model, and $\mathcal{P}^{\prime}$ by
that of the atomic limit (dashed and dotted curve of
fig.\ref{ter_cv4}).} \label{ter_kp1}
\end{figure}

\begin{figure}
\epsfig{file=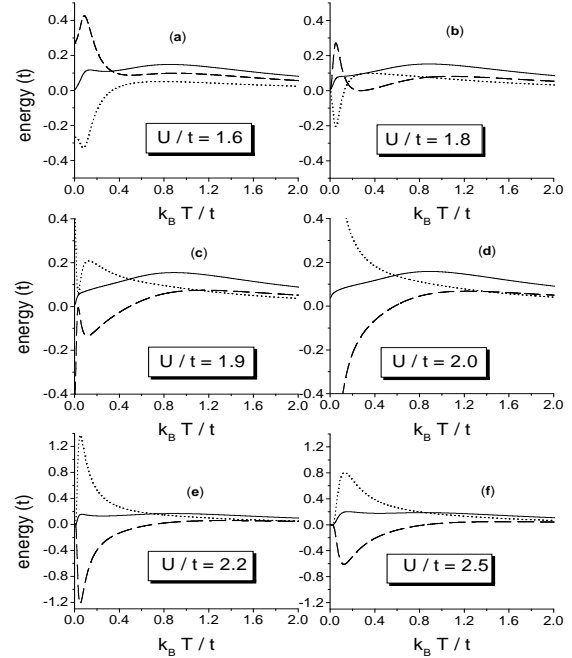,width=8.5cm,height=10cm,clip=}
\caption{The temperature dependence of the kinetic (dashed) and
potential (dotted) contributions to the specific heat (solid), at
half filling and different moderate values of $U$. Contrary to the
case $U\gg t$ of fig.\ref{ter_kp1}, at moderate coupling
$\mathcal{K}^{\prime}$ and $\mathcal{P}^{\prime}$ are competing,
since they have relatively large contributions of opposite signs
at roughly the same temperature. The peaks of $C_V$ are thus
'kinetic' (resp.'potential') when $\mathcal{K}^{\prime}$
(resp.$\mathcal{P}^{\prime}$) prevails on the other. Notice that
the low-temperature peak changes its origin from kinetic to
potential across the metal insulator transition point; the
high-temperature one, present up to $U/t \simeq 2.5$, is instead
always of kinetic origin. (For editing reasons the two bottom
figures have a different y-axis scale.)} \label{ter_kp2}
\end{figure}

In our model any peak of the specific heat has thus to be
ascribed just to `charge' excitations. We have seen in section
IV-D that, when varying the parameters $U$ and $\rho$, the peaks
can merge into one, and possibly reappear. In the following we
shall discuss such a structured behavior through the kinetic and
potential contributions to $C_V$, namely the derivatives
$\mathcal{K}'$ and $\mathcal{P}'$ with respect to the temperature
of $\mathcal{K}$ and $\mathcal{P}$, defined when
giving the internal energy (\ref{energy}).\\
We start by the case of the strong coupling ($U \gg t$), where
our results show a two-peak structure for
positive $U$ and $\rho<1$ (see fig.\ref{ter_cv4}-(b)), as well as
for negative $U$ at any filling (see fig.\ref{ter_cv7}-(b)).
Since in these regimes the characteristic energy scales of the
kinetic term ($t$) and the potential term ($U$) of the
Hamiltonian are well separated, it is expected that each of the
two peaks is related to one of these terms. In
fig.\ref{ter_kp1} we have thus plotted $\mathcal{K}^{\prime}$ and
$\mathcal{P}^{\prime}$ for $U/t=16$ and $\rho=0.75$: the two
peaks are indeed in perfect correspondence with the contributions
of $\mathcal{K}$ and $\mathcal{P}$. It is also worth stressing
that these two contributions can be quite well described at strong
coupling in terms of two different models: explicitly, the
low-temperature kinetic behavior is captured by the $U=\infty$
model for positive $U$'s (fig.\ref{ter_cv4}-(b)) and by the $XX0$
model for negative $U$'s (fig.\ref{ter_cv7}-(b)); the
high-temperature potential behavior is instead described by the
atomic limit.
In contrast, in the regime of moderate $U$'s the two energy scales become
comparable, and the above argument is not applicable. This gives
rise to a completely different scenario; for instance, at half-filling
we observe that by lowering $U$ the single strong-coupling peak
splits into two, whereas for $\rho<1$ the two strong-coupling peaks
merge into a single one.
In practice, while for $|U| \gg t$ the kinetic and potential
terms decouple, at moderate $U$'s it is the {\it competition}
between the two kinds of energy that determines the actual shape
of the specific heat.
\\This can be understood by recalling the structure of the energy
spectrum (see eq.(\ref{SPE}));
both terms can be expressed in terms of the quantum numbers
$n_k^A$, where the total number of $A$-sites is not a fixed
quantity, but can vary in the range $N_A \in [N/2 \, ; N]$ (the
electron number $N$ being obviously fixed). This property actually
yields the competition between
$\mathcal{P}$ and $\mathcal{K}$: indeed the kinetic term may
favor the decrease of $N_A$, in order to eliminate possible
positive contributions of $-2 t \cos{k}$, whereas the potential
term favors the increase of $N_A$ ({\it i.e.} the breaking of
on-site pairs). This competition is already active at
$T=0$, causing the appearance of the different regions in the
ground state phase diagram.
\\At finite temperature two more mechanisms enter driving such
competition: i) the density $\rho_A$ of $A$-carriers varies with
$T$, according also to the values of $U$ and $\rho$; ii) the
kinetic term exhibits the usual thermal fluctuations. The former
represent the crucial difference with respect to an ordinary free
spinless fermion model, where only thermal excitations are
present, at {\it fixed} number of carriers. Notice also that the
variability of $\rho_A$ can happen to contrast the effect of
thermal fluctuations: this is the case when $\rho_A$ decreases
with $T$, since this would yield a reduction of $\mathcal{K}$,
while thermal fluctuations would lead to an increase of it. As a
consequence, a further competition, concerned with the purely
kinetic contribution, may occur.\\
In fig.\ref{ter_kp2} we plot the derivatives
$\mathcal{K}^{\prime}$ and $\mathcal{P}^{\prime}$ of the kinetic
and the potential parts for various moderate $U$'s at
half-filling. Starting from $U/t=1.6$ we observe that at low
temperatures both $\mathcal{K}^{\prime}$ and
$\mathcal{P}^{\prime}$ exhibit a peak at nearly the same
temperature $T_1$; this is due to the fact that in this regime
they are driven by the same mechanism (formation of pairs from
singly occupied sites). The two contributions of opposite signs
do not completely cancel each other; the kinetic one prevailing, a
kinetic low-temperature peak appears in $C_V$. Notice that the
value of $C_V$ at the peak is relatively small with respect to
that of ${\mathcal{K}}^{\prime}$ and ${\mathcal{P}}^{\prime}$;
this is just the hallmark of a competition between the two
contributions. \\At a higher temperature $T \simeq T_2$, located
in between the two peaks of $C_V$, $\mathcal{K}'$ has a flat
minimum and $\mathcal{P}'$ a flat maximum. Finally, at still
higher values of temperature, $\mathcal{K}'$ exhibits a second
maximum at $T_3$, and $\mathcal{P}'$ is smoothly decreasing; in
correspondence, $C_V$ exhibits the second
peak, of kinetic origin.\\
As $U$ is increased (fig.(b)), the value of $T_{1}$ decreases and
the absolute height of both the above contributions drastically
vanishes, so that the low-temperature peak becomes a sort of
`shoulder'. At the same time, the minimum of the kinetic
contribution and the maximum of the potential contribution located
around $T_2$ have become more pronounced, and $T_2$ itself has
decreased (see fig.(c)). As $U$ reaches the value $2 t$ of the
metal-insulator transition, both $T_1$ and $T_2$ vanish and the
magnitude of the corresponding extrema become infinite (fig.(d)).
For $U> 2t$ (figs. (d) and (e)), the $T_2$ extrema are regularly
restored and, since the $T_1$-extrema have disappeared, they
become the new low-temperature extrema. At this temperature $C_V$
exhibits now a new peak. Thus, for $U>2 t$, the potential
contribution prevails on the kinetic one, and the nature of the
low-temperature peak changes with respect to the case $U<2t$.
Notice that $T_2$ now increases with $U$ (figs.(e) and (f)).
Finally, at higher temperatures another broad peak originates
from the (old) second maximum of the kinetic part. Such
high-temperature ($T_3$) peak is very broad, and it definitely
disappears when $U$ is further increased above $3t$.\\
The above observations show that at half-filling, passing through
the point $U=2t$ of the metal-insulator transition, the nature of
the low-temperature peak changes its origin from a kinetic to a
potential one, whereas at moderate $U$'s a further peak of kinetic
origin appears at higher temperatures. In  pass let us also
notice that at strong coupling a kinetic(potential) peak is a
peak to which only
${\mathcal{K}}^{\prime}$(${\mathcal{P}}^{\prime}$) basically
contributes, ${\mathcal{P}}^{\prime}$(${\mathcal{K}}^{\prime}$)
being almost vanishing (see fig.\ref{ter_kp1}); in contrast, at
moderate $U$'s a kinetic(potential) peak is a peak for which the
kinetic contribution slightly prevailing on the
potential(kinetic) one.

The results obtained for our model can be compared with those
concerning the ordinary Hubbard model. In the strong coupling
regime of this model the low-temperature peak is attributed to
spin excitations (the corresponding temperature being of the
order of $J=4 t^2/U$), whereas the high-temperature peak is
related to the charge excitations (since it is located at $k_B T
\sim U$). With lowering $U$, it is widely accepted that the two
peaks merge at $U \simeq 4 t$; however, some investigations have
been carried out at still lower $U$'s, showing that a double peak
structure reappears for D=1\cite{KUO} and D=2\cite{PASC,DUFFY}. It
is customary to relate the origin of these new peaks again to spin
and charge degrees of freedom respectively. \\The Hubbard model
is considered the paradigm within strongly correlated systems, so
that the presence of a two-peak structure in the specific heat of
such systems tends naturally to be interpreted as the signature of
spin and charge excitations.
\\However, in the authors' opinion, not enough attention has
been devoted to the effect that further interaction terms in the
Hamiltonian have on the specific heat. To this purpose, the exact
results obtained for our model show that, when a possible
competition between single and paired carriers is taken into
account, the specific heat turns out to exhibit a structured
two-peak behavior, in spite of the fact that only charge degrees
of freedom are involved. Although our model neglects some terms
such as nearest neighbors charge interaction ($ \sim V\hat{n}_{i
\sigma} \hat{n}_{j \sigma^{\prime}}$), we believe that it can
reproduce some features of realistic materials which are not
explicitly taken into account in the ordinary Hubbard model,
namely: a) the opening of the gap at a {\it finite} value of
$U/t$, {\it i.e.} at a finite value of pressure on the sample; b)
the lack of particle-hole symmetry, observed in heavy fermion
compounds; c) the presence of a mechanism favoring the kinetic of
paired carriers, as it is the case in cuprate superconductors.
\\In view of these observations, we suggest that the
interpretation of a two-peak structure in $C_V$ may not
necessarily be related to spin and charge excitations; a
comparison with the behavior of pure spin quantities, such as
magnetic susceptibility, in correspondence of the peaks
temperature would be more probative.
\section{Conclusions}
In this paper we have calculated the exact thermodynamics of an
Extended Hubbard model by means of Sutherland Species technique,
which we had previously introduced to determine the ground state
properties of the same model\cite{DOMO-RC}. The model describes a
competition between the dynamics of single carriers and that of
short-radius paired carriers; such competition is modulated by
the values of the electron filling $\rho$ and the on-site Coulomb
repulsion $U$. We have calculated the partition function of the
model, and derived the finite temperature behavior of different
physical quantities, namely the chemical potential, the
compressibility, the local magnetic moment, and the specific
heat. We have discussed the changes of such observables across
the point of the metal-insulator transition $U=2t$, providing
explicit low-temperature expressions for $C_V$ and $\mu$; in
particular $\mu$ is found to undergo an unusual transition from a
linear to a $T\ln T$ dependence. We have then focused on the
specific heat, which turns out to exhibit interesting features,
such as a nearly universal crossing point and a double peak
structure. The two peaks, which are shown to be related to charge
degrees of freedom only, are present in ranges of $U/t$ both
below and above the metal-insulator transition value. We have
discussed the two peaks in terms of kinetic and potential
contribution to the spectrum, outlining the differences between
the cases of strong coupling and moderate coupling, and comparing
our results with that of the ordinary Hubbard model.\\

The method presented here to derive the partition function of our
model can be applied, with straightforward generalization, to further
integrable extended Hubbard models \cite{DOMO-IJMB} involving two
Sutherland species. Work is in progress along these lines.

\end{multicols}

\begin{references}
\bibitem{NOTAUinf} In fact, it can easily be checked that the
Hamiltonian of the $U=\infty$ Hubbard model can be rewritten as
${\mathcal{P}} {\mathcal{H}}_t$, where ${\mathcal{P}}$ rules out
double occupancy.
\bibitem{NOTE} As to the value of $Y$, we observe that the models with
$Y > 0$ can be mapped into those with $Y<0$ through a mere redefinition
of the phase of the doubly occupied state, {\it i.e.}
$|\downarrow \uparrow \rangle_j \longrightarrow (-1)^j |\downarrow
\uparrow \rangle_j$, $j=1,\ldots L$. Therefore, the investigation
of the range $Y \le 0$ is completely general.
\bibitem{NOTAGP} Notice that, with respect to \cite{DOMO-RC}, here it
is sufficient to deal only with the off-diagonal part
$\hat{T}_{i,j}$ of the full generalized permutator $\Pi_{i,j}$
introduced there.
\end{references}
\end{document}